\documentclass{article}[12pt,a4paper,onesided]
\usepackage[margin=20mm]{geometry}
\usepackage{listings}
\usepackage{amsmath}
\usepackage{amsthm}
\usepackage{caption,subcaption}
\usepackage{array}
\usepackage{mdwmath}
\usepackage{multirow}
\usepackage{mdwtab}
\usepackage{eqparbox}
\usepackage{multicol}
\usepackage{amsfonts}
\usepackage{multirow,bigstrut,threeparttable}
\usepackage{amsthm}
\usepackage{array}
\usepackage{bbm}
\usepackage{epstopdf}
\usepackage{mdwmath}
\usepackage{mdwtab}
\usepackage{eqparbox}
\usepackage{tikz}
\usetikzlibrary{positioning, calc, shapes.geometric, shapes.multipart,
  shapes, arrows.meta, arrows,
  decorations.markings, external, trees, backgrounds,positioning,fit}
\usepackage{tikz-cd}

\usepackage{latexsym}
\usepackage{amssymb}
\usepackage{bm}
\usepackage{amssymb}
\usepackage{graphicx}
\usepackage{mathrsfs}
\usepackage{mathtools}
\usepackage{epsfig}
\usepackage{psfrag}
\usepackage{setspace}
\usepackage{algorithm}
\usepackage{algorithmic}
\usepackage{csquotes}

\usepackage[backend=bibtex,
citestyle=authoryear-comp,
bibstyle=authoryear-comp,
autocite=plain,
sorting=nyt,
hyperref=true,
maxcitenames=2,
maxbibnames=100,
isbn=false,
url=false,
doi=false]{biblatex}
\usepackage{hyperref}
\usepackage{cleveref}

\usepackage{soul}

\def\biomarker{S}
\def\biomarkerValue{s}
\def\CI{\text{CI}}

\usepackage{preamble}
\def\selfcontained{{\text{self-contained}~}}
\def\Selfcontained{{\text{Self-contained}~}}

\title{Selective randomization inference for subgroup effects with continuous biomarkers
}
\author{
Zijun Gao\footnote{\small Marshall School of Business, University of Southern California, USA}
}

\begin{document}
\maketitle



\begin{abstract}
    Randomization test is a popular method for testing causal effects in clinical trials with finite-sample validity.
    In the presence of heterogeneous treatment effects, it is often of interest to select a subgroup that benefits from the treatment, frequently by choosing a cutoff of a continuous biomarker.
    However, selecting the cutoff and testing the effect on the same data may fail to control the type I error.
    To address this, we propose to use {``self-contained''} methods for selecting biomarker-based subgroups (cutoffs) and use conditioning to construct valid randomization tests for the subgroup effect.
    Compared to sample-splitting-based randomization tests, our proposal is fully deterministic and uses up the entire selected subgroup for inference thus more powerful.
    Moreover, we demonstrate scenarios in which our procedure has comparable power to the randomization test with oracle knowledge of the benefiting subgroup.
    In addition, our procedure is as computationally efficient as standard randomization tests.
    Empirically, we illustrate the effectiveness of our method on simulated datasets and the German Breast Cancer Study.

\end{abstract}

\section{Introduction}

Understanding heterogeneous treatment effects across individuals has attracted attention in clinical trials with the rise of personalized medication \parencite{lesko2007personalized}.
When treatments are heterogeneous, the causal estimand of interest typically pertains to a subgroup rather than the full population \parencite{hernan2010hazards}.
Often, subgroups are specified by the measurements of biomarkers 
\parencite{cagney2018fda, food2012enrichment}.
As illustrated in our motivating dataset (\Cref{sec:GBCS}), a typical case 
is choosing a cutoff point of a continuous biomarker \parencite{Regan2011, Entresto2015, FDA2020, Janjigian2021}.

Although biomarker-driven subgroup analyses are popular, concerns about their statistical validity have emerged \parencite{mcshane2023finding, huber2025methodological}.
For example, model-based analyses, which rely on model assumptions and asymptotic approximations, can exhibit inflated type I errors in the presence of model misspecification and small-sample clinical trials \parencite{athey2015machine,burke2015three}.
In contrast, design-based methods \parencite{fisher_design_1935, pitman_significance_1937}, which leverage the randomness of treatment assignment for inference, offer finite-sample validity depending only on the randomization mechanism controlled by the trial experimenters.


However, most current design-based analyses are performed on either the full sample (``all-comers'') or pre-specified subgroups.
Applying randomization tests to data-driven subgroups selected from the same data often compromises the statistical validity.
To handle data-driven subgroups, a standard approach is to employ sample splitting
\parencite{cox_note_1975}.
Despite its convenience and flexibility, sample splitting incurs efficiency loss by using only part of the data for inference and yields non-deterministic results that depend on the realized data split.
A more recent work \parencite{freidling2024selective} employs conditional randomization tests to handle data-driven subgroups,
which performs well for selections based on discrete biomarkers but may exhibit reduced power and face computational challenges when applied to continuous biomarkers.
In light of the considerations, in this paper, we propose a method of selecting continuous biomarker cutoffs and constructing valid and powerful randomization-based tests for the selected subgroup.



\subsection{Motivating dataset:  German Breast Cancer Study Group trial}\label{sec:GBCS}

We first introduce the dataset that motivates our approach.
The German Breast Cancer Study Group (GBCS) Trial 2 \parencite{schumacher1994randomized, schmoor1996randomized, stallard2022adaptive} investigates the effect of hormone therapy on breast cancer recurrence time.
A total of $473$ patients\footnote{The trial enrolled $473$ participants; we focus on the $439$ patients with complete covariate information.} were enrolled and randomized, with a $40\%$ probability receiving the treatment.
Previous studies have shown that progesterone receptor count (biomarker) is associated with higher hormone absorption and larger treatment effect.
In light of this monotonic relationship, researchers are motivated to select a biomarker threshold and test the subgroup treatment effect associated.


\subsection{Proposal overview}\label{sec:overview}

\begin{figure}[tbp]
        \centering
        \begin{minipage}{1\textwidth}
        \centering
                \includegraphics[clip, trim = 0cm 0cm 8cm 0cm, height = 8cm]{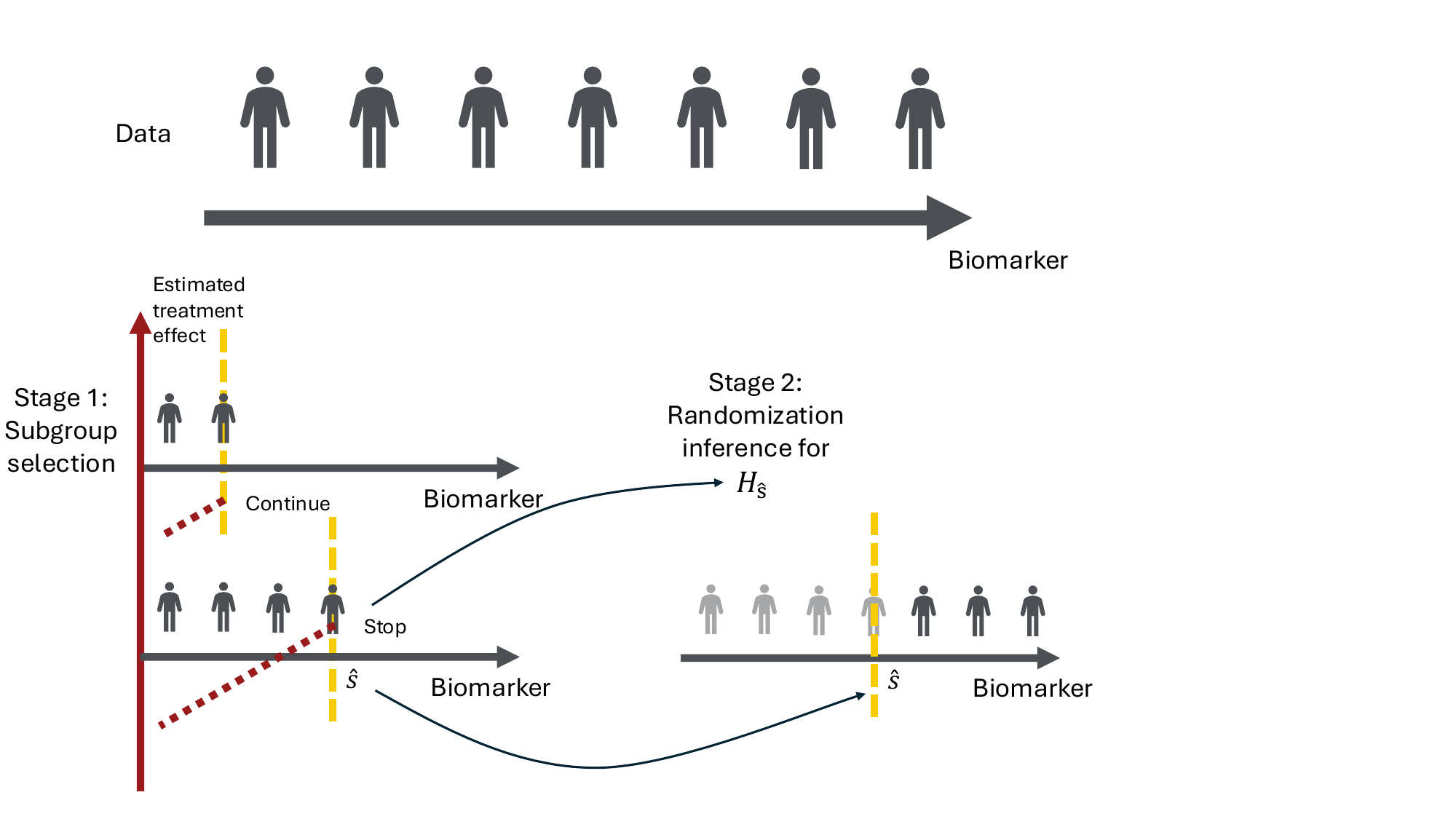}
        \end{minipage}
        \caption{Pipeline of the proposed method. 
        Units are ordered in ascending order regarding their biomarker values.
        In stage 1, we gradually reveal the outcomes of the units and stop once the estimated treatment effect at the current biomarker value, denoted by $\hat{s}$, exceeds zero.
        In stage 2, we perform randomization inference by reassigning the treatment of the selected group $\calS_{\hat{s}} = \{i: \biomarker_i > \hat{s}\}$ while keeping the treatment of the unselected group $\calS_{\hat{s}}^c$ fixed.}
        \label{fig:diagram}
\end{figure}



We provide an overview of our proposal (\Cref{fig:diagram}), which consists of a selection stage and an inference stage.
As in the GBCS data (\Cref{sec:GBCS}), we assume that larger biomarker values are associated with greater treatment effects.
\begin{enumerate}
    \item {\bf Selection} (Stage 1 of \Cref{fig:diagram}).
   We arrange the units in ascending order of their biomarker values.
    We reveal the treatment assignments sequentially, and estimate the treatment effect at each newly revealed biomarker value.
    We stop the process once the estimated treatment effect becomes positive\footnote{The stopping rule based on the estimated treatment effect becoming positive can be generalized to require the estimated value exceeding a prespecified threshold. }, and select the corresponding biomarker value as the cutoff.

    
    \item {\bf Inference} (Stage 2 of \Cref{fig:diagram}). 
   We apply the randomization test to the selected group (units whose biomarker values exceed the selected cutoff) while fixing the treatment assignments of the unselected units.
    In common experimental designs, this is equivalent to applying a standard unconditional randomization test to the selected subgroup.
\end{enumerate}
Our procedure is proven to achieve finite-sample validity without requiring correct modeling of outcomes, biomarkers, or unit dependencies.
We also show that the selected cutoff converges to the true threshold\footnote{The ``true threshold'' refers to the biomarker value above which units begin to benefit.} at the optimal rate\footnote{This rate is optimal in the problem of finding the zero point of a monotonic Lipschitz function.
} $n^{-1/3}$ ($n$ denotes the sample size).
In addition, our approach uses all the units in the selected subgroup, and achieves comparable power to the randomization test that knows the true benefiting subgroup in advance.
Finally, our procedure is fully deterministic and as computationally efficient as a standard (unconditional) randomization test.

We demonstrate the efficacy of our procedure on the GBCS dataset (\Cref{sec:GBCS}) in comparison with the sample splitting method and the Bonferroni correction method.
As shown in \Cref{tab:GBCS}, our approach selects the subgroup with at least one progesterone receptor, covering $85.6\%$ of the sample, and yields a p-value of \(1.6 \times 10^{-3}\) --- the only p-value significant at level $0.01$.
In contrast, the Bonferroni correction (based on $19$ pre-specified candidate cutoffs) selects the same subgroup but produces a p-value ten times larger.
The sample splitting approach (using a random 50-50 split) selects a higher cutoff of $16$ receptors, yielding a smaller subgroup covering only $62.4\%$ of the patients, and a large p-value despite the stronger treatment effect in its selected subgroup.

\begin{table}[tbp]
\centering
\caption{Comparison of selected subgroups and the associated p-values. We compare our procedure, the sample splitting method, and the Bonferroni correction method.  
Further details are provided in \Cref{sec:real.data}.
}
\label{tab:GBCS}
\begin{tabular}{c|ccc}
\toprule
Method & Selected subgroup & Selection rate & P-value \\
\midrule
Sample splitting        & $\geq 16$ receptors & 62.4\% & $40.88 \times 10^{-2}$ \\
Bonferroni  & $\geq 1$ receptors & 85.6\% & $1.71 \times 10^{-2}$\\
Proposal    & $\geq 1$ receptors & 85.6\% & 
$0.16 \times 10^{-2}$ \\
\bottomrule
\end{tabular}
\end{table}

\vspace{0.2cm}
\noindent\textbf{Organization}. 
In \Cref{sec:set.up}, we formulate the problem, introduce randomization inference and biomarker-based subgroups, and conclude with a review of related works.
In \Cref{sec:method}, we present our proposal in detail including the \selfcontained selection and the randomization tests afterwards.
In \Cref{sec:theory}, we establish the validity of the our proposed test and analyze the quality of the selected subgroup as well as the power of the test.
In \Cref{sec:simulation}, we evaluate the performance of our procedure through simulation studies.
In \Cref{sec:real.data}, we apply our method to the motivating GBCS dataset.
In \Cref{sec:discussion}, we conclude with discussions and future research directions.
Method extensions, proofs, and additional simulations are provided in the Appendix.

\vspace{0.2cm}
\noindent\textbf{Notations}.
For an integer \( n \), let \( [n] \) denote the set \( \{1, 2, \ldots, n\} \).
For a set \( \mathcal{S} \), we use $\calS^c$ to denotes its complement set, and $|\calS|$ to denote its cardinality.
For a subset \( \mathcal{S} \subseteq [n] \) and a vector \( Z \in \mathbb{R}^d \), we denote by \( Z_{\mathcal{S}} \) the sub-vector \( (Z_i)_{i \in \mathcal{S}} \), and denote by \( Z_{\mathcal{S}^c} \) the sub-vector \( (Z_i)_{i \notin \mathcal{S}} \).
We use the symbol without any subscript to denote the collection of values for the entire sample, such as $Z = (Z_1, \ldots, Z_n)$.
We use $\Phi(\cdot)$ to denote the CDF of the standard normal distribution.

\section{Problem formulation}\label{sec:set.up}

\subsection{Potential outcome model}\label{sec:potential.outcome.model}

We consider the potential outcome model \parencite{rubin1974estimating}.
Suppose there are $n$ units. 
For unit $i$, we use $Y_i(0)$ to denote the potential outcome under control and $Y_i(1)$ for the potential outcome under treatment.
Unit $i$ is also associated with a continuous\footnote{Our procedure also directly applies to the case where the biomarker is ordinal but not continuous.} biomarker $\biomarker_i$.
\footnote{Our methods can be directly generalized to incorporate covariates in addition to the biomarker.}We use $W_i = \{\biomarker_i, Y_i(0), Y_i(1)\}$ to denote the biomarker and potential outcomes. 
The treatment assignment indicator $Z_i \in \{0, 1\}$ specifies whether unit $i$ receives the treatment ($Z_i = 1$) or belongs to the control group ($Z_i = 0$). 
The potential outcome of the realized treatment, $Y_i = Y_i(Z_i)$, is observed.
See \Cref{tab:notations} for a summary.

\begin{table}
\begin{center}
\caption{Table of notations under the potential outcome model.}
\label{tab:notations}
\begin{tabular}{p{2cm}|p{10cm}}
    \toprule
    \textbf{Notation} & \textbf{Description} \\ \midrule
    $Z_i \in \{0, 1\}$           & Treatment assignment: $0$ for control, $1$ for treatment. \\ 
    $Y_i(0)$, $Y_i(1)$      &   Control and treatment potential outcomes. \\
    $Y_i$           & Observed outcome. \\ 
    $\biomarker_i$           & Biomarker value. \\ 
    $W_i$ & The biomarker and potential outcomes, that is, $(\biomarker_i, 
    Y_i(0), Y_i(1))$. \\ \bottomrule
\end{tabular}
\end{center}
\end{table}

\subsection{Randomization inference}\label{sec:randomization.inference}


In the seminal work \parencite{fisher_design_1935}, Fisher considers testing the sharp null hypothesis
\begin{align}\label{eq:hypothesis}
    H: Y_i(1) - Y_i(0) = 0, \quad i =1, \ldots, n.
\end{align}
Under the sharp null, all the potential outcomes can be determined based on the data observed, that is $Y_i(1) = Y_i(0) = Y_i$. 
Let $T = T(Z, W)$ be a test statistic depending on the observed treatment assignment $Z$ and $W$ including potential outcomes and the biomarker value.
We assume a larger test statistic value is associated with stronger evidence of rejecting the sharp null. 
In Fisher's randomization test, $W$ is regarded as fixed and the randomness comes exclusively from the treatment assignment $Z$. 
The distribution of the treatment assignment $Z$ is known, and as a result, the distribution of $T(Z, W)$ is accessible under the sharp null.
The randomization p-value is defined by comparing the observed test statistic to the reference null distribution,
\begin{align}\label{eq:pvalue.RT}
    P(Z, W):=\Prob^*\left(T(Z^*, W) \geq T(Z, W) \mid Z, W \right),
\end{align}
where $\Prob^*$ denotes the distribution of the alternative treatment assignment $Z^*$.
The finite-sample type I error is always controlled,
\begin{align}\label{eq:typeIError}
    \PP\left(P(Z, W) \le \alpha \mid W \right) \le \alpha, \quad \alpha \in [0,1],
\end{align}
despite of the experiment design, data generating mechanism of $(S_i, 
Y_i(0), Y_i(1))$, and the test statistic used.

Rather than testing the sharp null~\eqref{eq:hypothesis} for the entire sample, we aim to select a subgroup and test the null within the selected units.
Particularly, for a data-driven subgroup $\hat{\calS}$, the null of interest takes the form,
\begin{align}\label{eq:hypothesis.subgroup}
    H_{\hat{\calS}}: Y_i(1) - Y_i(0) = 0, \quad i \in \hat{\calS}.
\end{align}
This null is referred to as partially sharp because we can only impute the potential outcomes for units in the selected subgroup \( \hat{\calS} \), but not for the entire sample \parencite{zhang_2023_randomization_test}.
For the notion of validity, we aim to control the type I error further conditioning on the selection $\hat{\calS}$,
\begin{align}\label{eq:typeIError.subgroup}         
    \PP\left(P(Z, W) \le \alpha \mid W, \hat{\calS} \right) \le \alpha, \quad \alpha \in [0,1].
\end{align}
By the tower property, the control of \eqref{eq:typeIError.subgroup} implies that of \eqref{eq:typeIError}.
Controlling \eqref{eq:typeIError.subgroup}  allows us to treat the selected subgroup and hypothesis as fixed when interpreting the results and ensures type I error control among trials with the specific subgroup selected.
This is crucial in objectively evaluating the evidence in a randomized experiment from
a regulatory perspective.

\subsection{Benefitting subgroup}\label{sec:optimal.subgroup}

We use \( \mathcal{S}^* \) to denote the benefiting subgroup, i.e., all units with a positive treatment effect, 
\begin{align}\label{eq:optimal.subgroup}
    \mathcal{S}^* = \{i \in [n]: Y_i(1) - Y_i(0) > 0\}.    
\end{align}
We define the true cutoff $s^*$ as
\begin{align*}
    s^* = \sup \{ s_i : Y_i(1) - Y_i(0) \le 0 \}.
\end{align*}
If the treatment effect increases with the biomarker value, then the benefiting subgroup satisfies \( S^* = \{ i \in [n] : s_i > s^* \} \).

In this paper, our goal is to determine $s^*$ and thus \( \mathcal{S}^* \) from the data and test the associated null hypothesis as defined in \eqref{eq:hypothesis.subgroup}.
We remark that our proposal can be directly extended to target subgroups of the form \( \{i \in [n] : Y_i(1) - Y_i(0) > c\} \), for some pre-specified threshold \( c \).


\subsection{Literature}

\subsubsection{Model-based analysis of heterogeneous treatment effect}\label{sec:model.based.HTE.analysis}

Model-based analysis of treatment effects has gained significant attention, where such analysis relies on a super-population model for covariates, treatment, and response.
A large body of work focuses on conditional treatment effect (CATE) estimation, often incorporating machine learning tools as a sub-component \parencite{Athey2016, Foster2019, Semenova2021, Nie2021, Wager2018, Kunzel2019}. While estimation has seen substantial progress, conducting inference for CATE remains more challenging.
Bootstrap-based methods have been developed to achieve simultaneous coverage over all covariate values \parencite{chernozhukov2014anti, chernozhukov2015post, chernozhukov2020bootstrap}, but they require well-specified models with regularity conditions and may be conservative due to the stringent requirement of uniform coverage.

Among model-based research, a line of works particularly relevant to our work focuses on identifying subgroups that benefit from treatment and conducting inference therein.
A common approach selects subgroups by first estimating the CATE and then applying a predefined criterion, which introduces selection bias. 
One strategy to avoid the selection bias involves uniform convergence guarantees, ensuring that CATE estimates remain uniformly close to their true values across subgroups. However, these methods can be overly conservative due to strict coverage restriction.
A simple and widely used approach is sample splitting, where data is divided into separate folds for subgroup selection and inference \parencite{box1957, Athey2016}. However, sample splitting uses only part of the data for inference and is susceptible to efficiency loss, and introduces additional randomness to its results. 
Performing multiple sample splits and aggregating results can reduce this randomness, though challenges remain, such as dependence across splits and the difficulty in merging multiple selected subgroups. 
Alternatively, when the subgroup is chosen from a pre-specified set, the inference problem can be framed within the multiple hypothesis testing framework \parencite{stallard2022adaptive}. However, the power of multiple testing procedures is less clear.

\subsubsection{Biomarker-based subgroup analysis}\label{sec:biomarker.subgroup.analysis}
Another line of related work focuses on using biomarkers to assist with subgroup identification and inference.
For binary or discrete biomarkers, relevant methods include \parencite{Wang2007, Spiessens2010, Jenkins2011, Friede2012, Stallard2014, Rosenblum2016, lai2019adaptive, Chiu2018}. 
For continuous biomarkers, it is typically believed that there is a monotone dependence of the treatment effects on the biomarker values \parencite{mcshane_finding_2023}.
Accordingly, the selected subgroup often consists of patients whose biomarker values exceed (stay below) a certain threshold.
Related approaches can be found in \parencite{stallard2014adaptive, lin_inference_2021}.
Among the above methods, a significant proportion are model-based, meaning their validity relies on correctly specifying models for the outcome, covariates, and dependencies among units. 
In contrast, we conduct biomarker-based subgroup analysis using a design-based approach.

\subsubsection{Selective inference}\label{sec:selective.inference}

A multitude of methods have been proposed to perform valid inference after selection for model-based inference, see \cite{kuchibhotla_post-selection_2022} for a comprehensive review.
A thread of works frames the valid post-selection inference as a simultaneous inference problem and controls the relevant error for all selections belonging to a pre-specified set \parencite{berkValidPostselectionInference2013, bachoc2020uniformly}.
However, the simultaneous control approach can be conservative since the coverage guarantee is provided for all models but is required only for one selected model.
Another thread of works, referred to as the conditional selective inference approach, requires appropriate error control conditional on the selection \parencite{lee_exact_2016, tian_selective_2018},
and relies on approximating the conditional distribution of the test statistic.
This thread relies crucially on the data generating mechanisms, test statistics, and selection methods, which may limit their applicability.
Our approach departs from the predominantly model-based literature by adopting a randomization-based framework.

\section{Method}\label{sec:method}


In this section, we discuss \selfcontained subgroup selection  in \Cref{sec:subgroup.selection} and the subsequent randomization inference in \Cref{sec:CRT}.

\subsection{Subgroup selection}\label{sec:subgroup.selection}

We first introduce the \selfcontained property of a selection method, which requires that the selected subgroup be fully determined by the treatment assignments of the unselected units.
Let $\calW$ denote the space of $W$.
\begin{definition}[\Selfcontained selection]\label{defi:self.contained}
   A selection mapping $\SS(\cdot; \cdot): \{0,1\}^n \times \calW \to 2^{[n]}$, $(z, w) \mapsto \SS(z; w)$ is called \selfcontained if and only if for any $w$, any $z$, $z'$ satisfying $z'_i = z_i$ for all $i \in  \SS^c(z; w)$,
    \begin{align*}
        \SS(z; w)
        = \SS(z'; w).
    \end{align*}
\end{definition}


\begin{remark}[Random \selfcontained selection]
    \Cref{defi:self.contained} can be further extended to random selection mappings, where the selection mapping depends not only on the covariates $w$ and treatment assignments $z$, but also external randomness.
    For example, sample-splitting-based selection is a random \selfcontained mapping.
    The type I error control in \Cref{prop:typeIError} for deterministic \selfcontained selections also holds for random \selfcontained selections.
    However, since random selection may produce irreproducible subgroup selection and encourage the manipulation of random seeds, we focus on deterministic \selfcontained selections in this paper.
\end{remark}

Next, we describe a family of selection methods satisfying the \selfcontained property in \Cref{defi:self.contained}. The selection methods are suitable when the treatment effect is expected to increase\footnote{In \Cref{appe:sec:extension} of the Appendix, we explore \selfcontained~selections for cases where the treatment effect first increases and then decreases with the biomarker value, as well as for settings involving multiple biomarkers.
} with the biomarker value.
\begin{itemize}
    \item {Step 1}. Among units with concealed treatment assignments, reveal the treatment assignment of the unit with the smallest biomarker value.

    \item {Step 2}. Evaluate the treatment effect at the current biomarker value.

    \item Alternate between Step 1 and Step 2 until the treatment effect becomes positive.
    Define the selected cutoff $\hat{s}$ as the current biomarker value, and the selected subgroup as $\mathcal{S}_{\hat{s}} := \{i : S_i > \hat{s}\}$.

\end{itemize}
See \Cref{fig:diagram} for an illustration and \Cref{algo:1} for details.
This family of selections is \selfcontained because 
the treatment assignments of the selected units with large biomarker values remain concealed throughout.
The stopping rule is designed to ensure that all screened units have non-positive treatment effects.
Moreover, if the treatment effect increases with the biomarker value, the selected subgroup consists of all units with positive treatment effects, effectively uncovering the target subgroup $S^*$.

\begin{example}\label{exam:selection.example}
We provide a concrete realization from the above family of \selfcontained selections by specifying how to evaluate the treatment effect at the current biomarker value and how to determine whether the effect is positive (the stopping rule).
To start, rank units in ascending order of their biomarker values and partition the sample into $n^{1/3}$ equally sized batches $\calB_j$.
At each step, reveal the treatment assignments for the next batch $\calB_j$ and estimate\footnote{The treatment effect estimator~\eqref{eq:diff.in.means} can be improved, especially when there are covariates $X_i$. 
For example, we can regress \( Y \) on \( X \) and \( \biomarker \) to estimate the marginal mean \( \EE[Y \mid \biomarker, X] \), as advocated in \cite{rosenbaum2002covariance}.The choice of regressor is flexible, such as methods based on linear models, tree-based methods, or neural networks, as long as the treatment assignment is excluded. We then use the residuals \( R_i \) to estimate the treatment effect, which is less variable than $Y_i$ since part of the variation due to the covariates and the biomarker reflected in \( \EE[Y \mid \biomarker, X] \) has already been removed from the outcome. } the treatment effect at the largest biomarker value of the batch by
\begin{align}\label{eq:diff.in.means}
    \sum_{i \in \calB_j} \frac{Z_i Y_i}{e_i} - \sum_{i \in \calB_j} \frac{(1-Z_i) Y_i}{1 - e_i}.
\end{align}
Here $e_i$ denote the probability of the $i$-th unit being treated, i.e., its propensity score.
Stop if the estimated treatment effect turns positive.
\end{example}

\begin{algorithm}[t]
\caption{Subgroup selection and selective randomization inference with a continuous biomarker}
\label{algo:1}
\begin{algorithmic}
\STATE
 \textbf{Input}: Data $(\biomarker_i, Z_i, Y_i)$, test statistic function $T(z, w)$; treatment assignment mechanism $\pi$; number of permutations $M$.\\
 \STATE \textbf{Stage 1: Subgroup selection.} 
 
 Order units increasingly in $\biomarker_i$. 
 Split the units according to $\biomarker_i$ into $n^{1/3}$ equal-sized batches $\mathcal{B}_j$.
  \FOR{$j = 1$ to $n^{1/3}$}
    \STATE Compute the estimated treatment effect $\hat{\tau}_j$ as in Eq.~\eqref{eq:diff.in.means}.
    \STATE If $\hat{\tau}_j > 0$, let $\hat{s} = \max\{S_i: i \in \calB_j\}$ and break.
\ENDFOR
\STATE \textbf{Stage 2: Selective randomization inference}.

Impute potential outcomes under $H_{\calS_{\hat{s}}}$. 
\FOR{t $= 1: M$}
\STATE Generate a new treatment assignment $\treatment^{*(t)}$ following the conditional distribution $\pi\left(\cdot \mid \treatment_{\calS_{\hat{s}}^c}^{*(t)} = \treatment_{\calS_{\hat{s}}^c}\right)$.
\ENDFOR
\STATE Compute $\hat{P}_M = \left({1+\sum_{t=1}^{M} \1\left\{T(\treatment^{*(t)}, \dataConditionedOn) \leq T(\treatment, \dataConditionedOn)\right\}}\right)/({1+M})$. 
\STATE \textbf{Output}: $\hat{s}$, $\calS_{\hat{s}}$, $\hat{P}_M$.
\end{algorithmic}
\end{algorithm}

\subsection{Conditional randomization test}\label{sec:CRT}

Provided with a selected subgroup $\calS_{\hat{s}}$, we test the associated partially sharp null $H_{\calS_{\hat{s}}}$.
For this purpose, we propose to use the conditional randomization p-value
\begin{align}\label{eq:pvalue.CRT}
    P(Z, W):=\Prob^*\left(T(Z^*, W) \leq T(Z, W) \mid Z, W, Z^*_{\calS_{\hat{s}}^c} = Z_{\calS_{\hat{s}}^c}\right).
\end{align}
Here $\Prob^*$ denotes the distribution of the treatment assignment with the treatment of the units not selected fixed at their observed values, i.e.,  $Z^*_{\calS_{\hat{s}}^c} = Z_{\calS_{\hat{s}}^c}$.
In other words, we only randomize the treatments of the selected units and leave the treatments of the unselected units unchanged.
See \Cref{exam:Bernoulli}, \Cref{exam:CRD}, \Cref{exam:stratified.CRD} for concrete examples of the conditional distribution.

The conditioning event \( Z^*_{\mathcal{S}_{\hat{s}}^c} = Z_{\mathcal{S}_{\hat{s}}^c} \) is both necessary and sufficient for the computability of the $p$-value in Eq.~\eqref{eq:pvalue.CRT} \parencite{zhang_2023_randomization_test}. 
On one hand, without this conditioning, it is possible that \(Z_i^* \neq Z_i\) for some \(i \notin \mathcal{S}_{\hat{s}}\), and \(T(Z^*, W)\) becomes uncomputable because \(Y_i(Z_i^*)\) is unknown under the partially sharp null; on the other hand, by fixing the treatment assignments for all units outside the selected subgroup, we only randomize the treatments within \(\calS_{\hat{s}}\) whose counterfactual outcomes are known under $H_{\calS_{\hat{s}}}$.


We can construct confidence interval for treatment effect of the selected subgroup by the duality between hypothesis testing and confidence set.
Explicitly, given a selected subgroup $\hat{\calS}$, we compute the p-value $P_c(Z, W)$ for a series of hypotheses indexed by $c$,
\begin{align*}
    H_{\hat{s}, c}: Y_i(1) - Y_i(0) = c, \quad i \in \calS_{\hat{s}}.
\end{align*}
We construct the confidence set for the treatment effect of the selected subgroup as
\begin{align}\label{rmk:confidenceSet}
    \CI_{\hat{\calS}}(\alpha) := \{c: P_c(Z, W) \ge \alpha\}.
\end{align} 

\begin{example}[Bernoulli design]\label{exam:Bernoulli}
    Consider a Bernoulli design where each unit receives treatment independently with a known probability \( e_i \), which may vary across units.  
    After conditioning on the treatment assignments of the units not in the selected subgroup, the units in the selected subgroup \(\hat{\mathcal{S}}\) still follow an independent Bernoulli design with the same treatment probabilities \( e_i \), \( i \in \hat{\mathcal{S}} \).
\end{example}

\begin{example}[Completely randomized design (CRD)]\label{exam:CRD}
    In a completely randomized design with \(n_1\) treated units and \(n_0\) control units, suppose we fix the treatment assignments for a subset of units containing \(n_1'\) treated and \(n_0'\) control. The treatment assignment for the remaining units then follows a new completely randomized design, assigning \(n_1 - n_1'\) to treatment and \(n_0 - n_0'\) to control among the remaining units.
\end{example}

\begin{example}[Stratified completely randomized design]\label{exam:stratified.CRD}
    A stratified completely randomized design first divides the sample into strata based on covariates, then performs independent completely randomized designs within each stratum. 
    After conditioning on the treatment assignments of units outside the selected group, the design is equivalent to independently applying CRD within each stratum conditional on the fixed assignments in that stratum (as illustrated in \Cref{exam:CRD}).
\end{example}

\section{Theoretical Properties}\label{sec:theory}

In this section, we present the conditional type I error control, characterize the quality of the selected subgroup (cutoff), and provide the power analysis of the proposed method.

\subsection{Validity}\label{sec:typeIError}


\begin{proposition}[Conditional type I error control]\label{prop:typeIError}
    For any \selfcontained selection satisfying \eqref{defi:self.contained}, the p-value in \eqref{eq:pvalue.CRT} controls the conditional type I error,
    \begin{align*}
        \Prob\left(\,P(Z, W) \leq \alpha \,\vert\, W, \calS_{\hat{s}}\right) \leq \alpha,\quad \forall\, ~ \alpha \in [0,1].
    \end{align*}
    Consequently, the test also controls the marginal type I error,
    \begin{align*}
        \Prob(\,P(Z,W) \leq \alpha ) \leq \alpha,\quad \forall\, \alpha \in [0,1].
    \end{align*}
\end{proposition}

We provide a detailed proof of \Cref{prop:typeIError} in the Appendix.  
As a direct corollary of \Cref{prop:typeIError}, we obtain a conditional coverage guarantee for the constructed confidence interval.
\begin{corollary}\label{coro:confidenceInterval}
    \Cref{prop:typeIError} implies that the confidence set at level $\alpha$ in Eq.~\eqref{rmk:confidenceSet}, denoted by $\calI_\alpha$, achieves the desired conditional coverage,
    \begin{align*}
        \PP\left(c^* \in \calI \mid W, \calS_{\hat{s}} = s\right) \ge 1 - \alpha, \quad \forall~s \in \calS_\tau(W),
    \end{align*}
    where $c^*$ is the true constant treatment effect.
\end{corollary}

\subsection{Power analysis}\label{sec:power}

Our randomization test is equivalent to applying a randomization test to \(\hat{S}\) as if the selected subgroup was pre-specified.
Thus, the power loss of our procedure compared to a randomization test with oracle knowledge of the true benefiting subgroup \(S^*\) reduces to the gap between \(\hat{S}\) and \(S^*\).

Below we focus on the selected subgroup \(\hat{S}\) in \Cref{exam:selection.example}. We bound the difference between \(\hat{\mathcal{S}}\) and the target \(\mathcal{S}^*\) under the following super-population model,
\begin{align}\label{eq:population}
    \begin{split}
        \biomarker_i &\sim \PP_{\biomarker}, \\
        W_i &\sim \text{Ber}(e(S_i)), \\
        Y_i(0) &= \mu_0(\biomarker_i) + \varepsilon_{i0}, \\
        Y_i(1) &= \mu_1(\biomarker_i) + \varepsilon_{i1}.
    \end{split}
\end{align} 
Here the biomarker is drawn from an unknown distribution $\PP_{\biomarker}$, the treatment assignment follows independent Bernoulli random variable with probability parameter $e(S_i)$, $\mu_0(\cdot)$, $\mu_1(\cdot)$ denote the conditional mean function of the control and treatment group, respectively, and $\varepsilon_{i1}$, $\varepsilon_{i0}$ denote zero mean random errors independent of $\biomarker_i$.
The treatment effect conditional on the biomarker is given by $ \tau(\biomarkerValue) := \EE[Y_i(1) -  Y_i(0) \mid \biomarker_i = \biomarkerValue] = \mu_1(\biomarkerValue) - \mu_0(\biomarkerValue)$.
With a slight abuse of notation, we let \( s^* := \sup\{s : \tau(s) \le 0\} \).

\begin{proposition}\label{prop:power}
Under the super-population model \eqref{eq:population}, and we further make the following assumptions.
\begin{enumerate}
    \item There exists a constant \( \underline{e} > 0 \) such that $
    \underline{e} \leq e(s) \leq 1 - \underline{e}$.  
    \item There exists \( B > 0 \) such that \( |Y_i(0)|, |Y_i(1)| \le B\).

    \item There exists a neighborhood \( \mathcal{N}(s^*) \) of the true cutoff $s^*$ satisfying the following properties.
    \begin{enumerate}
        \item There exists $\delta > 0$ such that for any \( s \notin \mathcal{N} \), if $s < s^*$, then \( \tau(s) < -\delta \); if $s > s^*$, then \( \tau(s) > \delta \).
        \item There exists $\underline{L}>0$ such that for any \( s', s \in \mathcal{N} \), \( s' < s \),
        \[
        \tau(s) - \tau(s') \geq \underline{L} (s - s').
        \]
        \item There exists $\underline{f} > 0$ such that the biomarker \(S\) has a density lower bounded by \(\underline{f}\) in \(\mathcal{N}(s^*)\).

        \end{enumerate}
\end{enumerate}
   Then for $n$ sufficiently large, the estimated cutoff $\hat{s}$ satisfies
    \begin{align*}
        \PP\left(\left|\hat{s} - {s}^* \right| \le C_1 \frac{\sqrt{\log(n)}}{n^{1/3}} \right) 
        \ge 1 - n^{-1/6} - e^{-C_2n^{1/3}},
    \end{align*}
    for some $C_1$, $C_2>0$ depending on $B$, $\underline{e}$, $\underline{L}$, $\underline{f}$.
\end{proposition}

The proof can be found in \Cref{appe:sec:proofs}.
As shown by \cite{kiefer1982optimum}, the minimax rate for estimating a monotone, Lipschitz continuous function is \( n^{-1/3} \).
Combined with Assumption 3(b) in \Cref{prop:power}, the zero point can only be estimated at the rate \( n^{-1/3} \).
Therefore, the rate achieved by our selected cutoff in \Cref{prop:power} is optimal up to a logarithmic factor of  the sample size.

\section{Empirical studies}

\subsection{Simulation}\label{sec:simulation}

We compare four randomization-based methods including the oracle one.
All methods adopt the \eqref{eq:diff.in.means} as the test statistic.
\begin{itemize}
    
    \item Randomization test with sample splitting (Sample splitting).
    We randomly divide the sample into two equal folds: on the selection fold, we compute a non-decreasing treatment effect estimator\footnote{On the selection fold, we estimate \(\mu_0(\biomarkerValue)\) and \(\mu_1(\biomarkerValue)\) using gradient boosting, and use their difference as the initial estimator $\tilde{\tau}(\biomarkerValue)$.
    We further let $\hat{\tau}(\biomarkerValue) \leftarrow \sup\{\tilde{\tau}(s') : s' < \biomarkerValue\}$ such that $\hat{\tau}(\biomarkerValue) $ is nondecreasing.} \(\hat{\tau}(\biomarkerValue)\) and let $\hat{\biomarkerValue} = \sup\{\biomarker_i: \hat{\tau}(\biomarker_i) \le 0\}$; on the inference fold, we apply a standard randomization test to the units whose biomarker value exceeds \(\hat{s}\).
    
    \item Randomization test with Bonferroni correction (Bonferroni).
    We specify a set of candidate cutoffs of the biomarker value (\(5k\%\),  \(k \in [20]\) quantiles of the biomarker value), and perform a standard randomization test to each candidate cutoff. We then apply Bonferroni correction to the resulting p-values and select the largest subgroup with a significant corrected p-value.

    \item Proposed approach (ART). We apply \Cref{algo:1} with a batch size of $20$.

     \item  Oracle randomization test (Oracle). The oracle method has access to the benefiting subgroup \(S_{\lambda^*}\) and applies a standard randomization test to \(S_{\lambda^*}\).
    
\end{itemize}
To make comparison fair, all randomization tests adopt the same test statistic \eqref{eq:diff.in.means}.
Randomization p-values are calculated using $M =  200$ Monte Carlo treatment reassignments.
The significance level is set at $\alpha = 0.05$.

We describe the default data-generating mechanism. We generate a dataset of $400$ units following the super-population model~\eqref{eq:population}. For each unit $i$, the biomarker $\biomarker_i$ is drawn from $\mathcal{N}(0, 4)$.
The treatment assignment $Z_i$ is a Bernoulli random variable with success probability $0.2$. 
The outcome $Y_i$ is generated according to
$Y_i = \mu_0(\biomarker_i) + Z_i \tau(\biomarker_i) + \varepsilon_i$, 
where $\mu_0(z) = z + z^2$ and $\varepsilon_i$ follows $\mathcal{N}(0, 16)$. For $\tau(z)$, we consider two options: a linear form $\tau(z) = \delta z$, and a sigmoid form $\tau(z) = 2\delta \frac{\exp(\delta z)}{1 + \exp(\delta z)} - \delta$, where the default effect size is $\delta = 6$. 
In the Appendix \Cref{appe:sec:simulation}, we explore a piecewise constant form of $\tau$.

\begin{figure}[tbp]
        \centering
        \begin{minipage}{0.35\textwidth}
        \centering
    \includegraphics[clip, trim = 0cm 0cm 6cm 0cm, height = 0.9\textwidth]{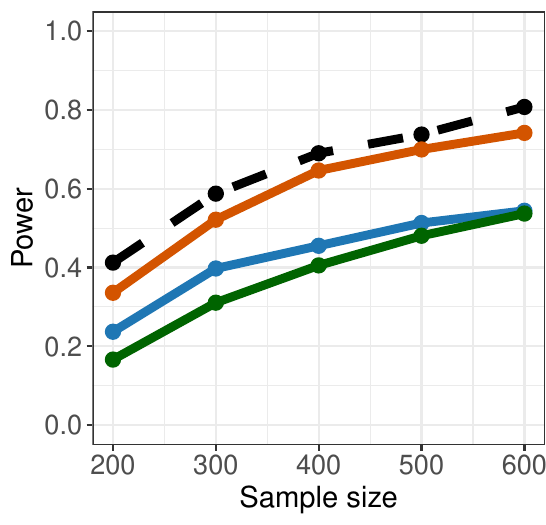}
        \end{minipage}
   \hspace{0.5cm}
   \begin{minipage}{0.35\textwidth}
        \centering
    \includegraphics[clip, trim = 0cm 0cm 0cm 0cm, height = 0.9\textwidth]{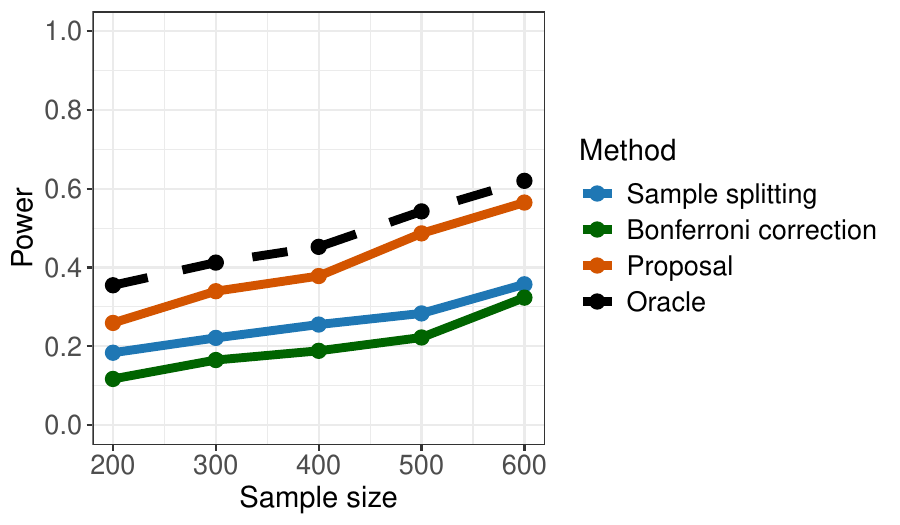}
        \end{minipage}\\         \vspace{0.2cm}
        \begin{minipage}{0.35\textwidth}
        \centering
            \includegraphics[clip, trim = 0cm 0cm 6cm 0cm, height = 0.9\textwidth]{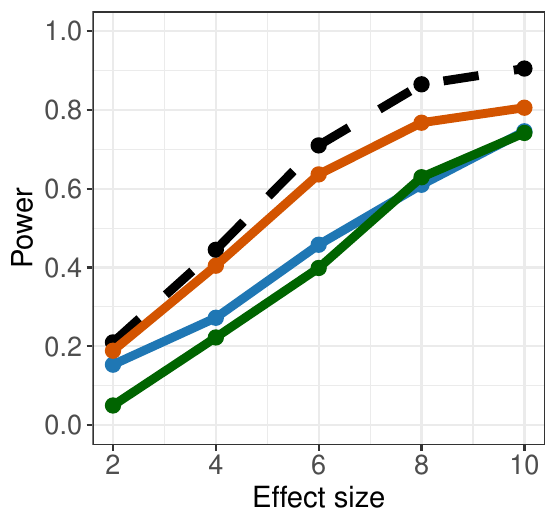}
        \end{minipage}
       \hspace{0.5cm}
        \begin{minipage}{0.35\textwidth}
        \centering
            \includegraphics[clip, trim = 0cm 0cm 0cm 0cm, height = 0.9\textwidth]{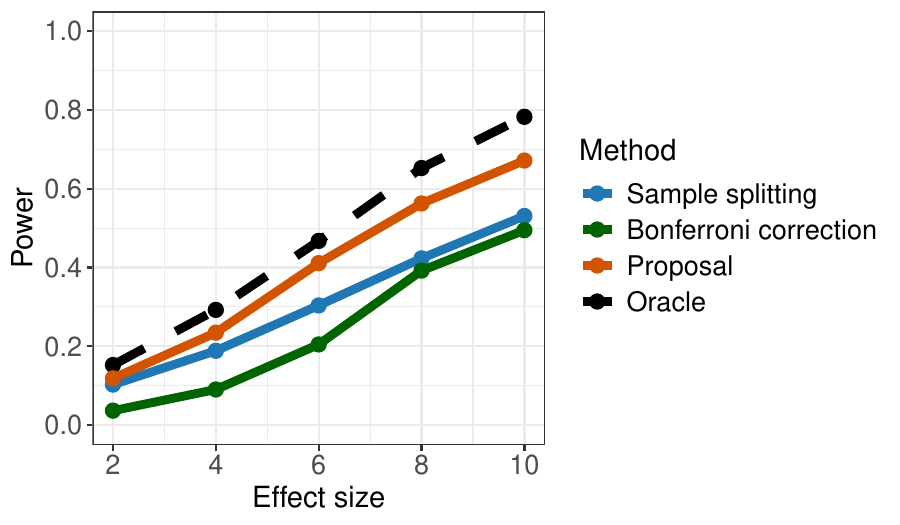}
        \end{minipage}\\
        \vspace{0.2cm}
         \qquad \text{(a) Linear $\tau$} \qquad \qquad \qquad \qquad \qquad \qquad     
        \text{(b) Sigmoid $\tau$}
\caption{Power comparison of four randomization-based methods. 
}
        \label{fig:power}
\end{figure}

We report power defined as the expectation of the proportion of the benefiting subgroup selected, multiplied by the indicator that $H_{\hat{\mathcal{S}}}$ is rejected, i.e.,
\begin{align*}
    \text{Power} = \EE\left[\frac{|\calS^* \cap \hat{\calS}|}{|\calS^*|} \cdot \1_{\{H_{\hat{\calS}}~\text{is rejected}\}}\right].    
\end{align*}
In the first row of \Cref{fig:power}, we vary the sample size from $n = 200$ to $600$; in the second row of \Cref{fig:power}, we vary the effect size from $\delta = 2$ to $12$.
We repeat each setting $400$ times and aggregate the results.



As shown in \Cref{fig:power}, our method achieves power comparable to the oracle method and outperforms both the sample splitting method and the Bonferroni correction method. 
The power improvement is robust across options of $\tau$, effect sizes, and sample sizes. 
In particular, the gap between the sample splitting method and our proposal becomes more pronounced with larger sample sizes, where a larger number of units are discarded by the sample splitting method.

\subsection{Real data analysis}\label{sec:real.data}

We expand the empirical study on the GBCS data in \Cref{sec:overview}. 
As described in \Cref{sec:GBCS}, the GBCS dataset is collected from a Bernoulli-design clinical trial with breast cancer recurrence time as the primary outcome. 
A summary table of the treatment and control group characteristics is provided in \Cref{tab:data} in the Appendix.
We aim to select a cutoff point for the progesterone receptor count (biomarker) and test the associated subgroup effect.

We compare the three randomization-based methods described in \Cref{sec:simulation}, 
except the oracle approach.
In all three randomization-based tests, we use the estimated coefficient from the Cox regression of recurrence time over hormone treatment as the test statistic.
In the Bonferroni correction method, we consider $19$ candidate subgroups $\{i: \biomarker_i \ge 1+c\}$, 
with $c \in \{-1, 0, 1, 3, 6, 
10, 15, 20, 25, 30, 
45, 60, 80, 100, 130, 
160, 200, 250, 400\}$. 
In our proposal (\Cref{algo:1}), we stop at the first batch $\calB_j$ where $1 - \Phi(\sqrt{|\calB_j|}\hat{\tau}_j / \hat{\sigma}_j)$ falls below $0.1$ (here $\hat{\tau}_j$ is the difference in means defined in \eqref{eq:diff.in.means}, and $\hat{\sigma}_j$ is its empirical standard deviation).
All other aspects of the methods in comparison
follow \Cref{sec:simulation}.

As shown in \Cref{tab:GBCS}, our approach selects the subgroup with at least one progesterone receptor and achieves the most significant $p$-value. 
Excluding patients with no receptor is biologically sensible, as such patients can't absorb the hormone and thus unlikely to benefit from the treatment.
The Bonferroni correction method selects the same subgroup, but the $p$-value is less significant due to the multiplicity adjustment for the $19$ hypotheses tested. 
When more candidate subgroups are tested, the adjustment will be more severe and the p-value will be further inflated.
For the sample splitting method, the selected subgroup is over $20\%$ smaller than that chosen by our proposal, and the associated $p$-value is less significant.
A smaller selected subgroup is often less desirable in practice, 
as trials aim to identify a large benefiting population to maximize the treatment's clinical impact.


\section{Discussion}\label{sec:discussion}

In this work, we propose a design-based test for treatment effects in data-driven subgroups defined by continuous biomarkers.
We propose using self-contained selections and performing a conditional randomization test within the selected subgroup.
Our proposal offers finite-sample type I error control without requiring a well-specified model, and achieves power comparable to a randomization test with oracle knowledge of the benefiting subgroup.
When applied to simulated and real datasets, our method produces desirable subgroups as well as significant $p$-values for the subgroup effect.

We outline several directions for future research.
First, more complex designs beyond the common assignment mechanisms \Cref{exam:Bernoulli}, \Cref{exam:CRD}, and \Cref{exam:stratified.CRD} have been adopted in practice. 
It is of interest to characterize the distribution of treatment assignments conditional on the assignments of a subset of units and to develop efficient methods for sampling from this conditional distribution.
Second, in certain biological applications, natural biomarkers may not exist, and researchers may instead learn a data-driven biomarker, such as a genetic risk score \parencite{marston_predicting_2020}, for subsequent subgroup selection.
It is of interest to learn how to construct a data-driven biomarker that is predictive while enabling valid testing of subgroup effects afterwards.
We provide an initial attempt in \Cref{appe:sec:extension} in the Appendix.
Third, this paper focuses on the  (partially) sharp null hypothesis which specifies the treatment effect for every individual. 
Recent research has explored the use of randomization tests for one-sided null hypotheses \textcite{caughey_randomisation_2023} and weak null hypotheses concerning average treatment effects \parencite{ding_paradox_2017, wu_randomization_2021, cohen_gaussian_2022}. It is of interest whether our approach could be extended to address these null hypotheses.

\section*{Acknowledgment}
We thank Qingyuan Zhao and Tobias Fielding for their valuable discussions. 
We thank Moulinath Banerjee for insightful comments on adaptively collected trial data.

\printbibliography

\appendix

\section{Method extensions}\label{appe:sec:extension.general}
\subsection{Non-monotone biomarkers}\label{appe:sec:extension}

\begin{figure}[tbp]
        \centering
        \begin{minipage}{0.4\textwidth}
        \centering
            \includegraphics[clip, trim = 0cm 0cm 22cm 0cm, height = 8cm]{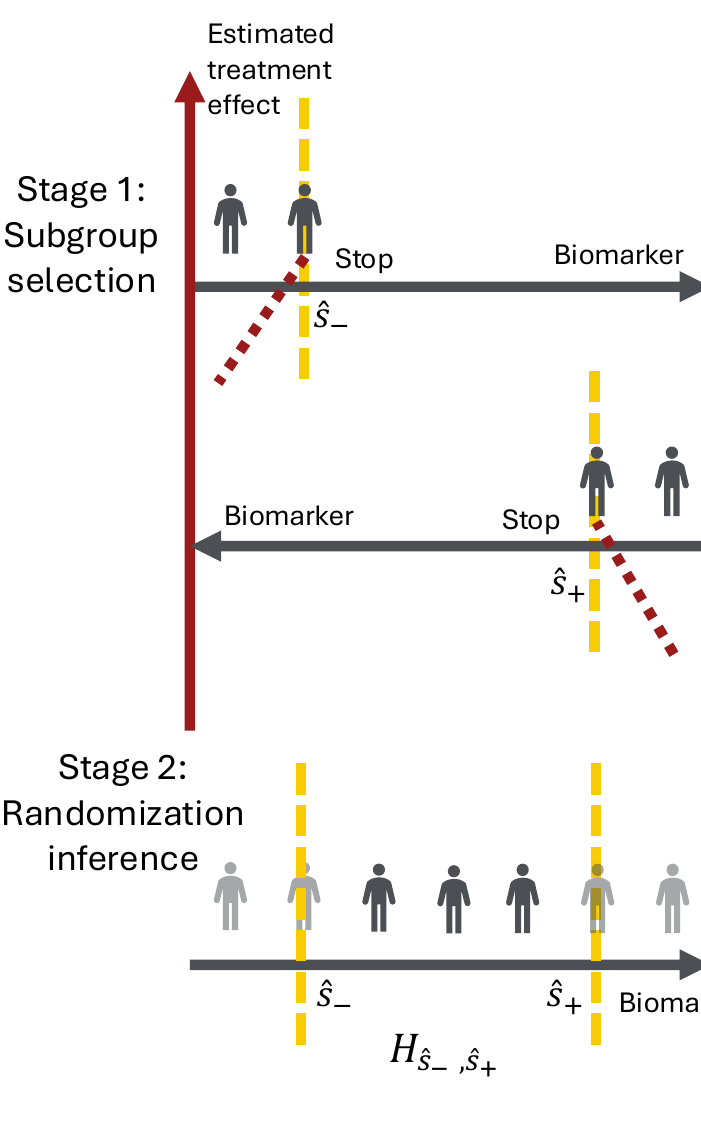}
            \subcaption{Arc-shape dependence}
        \end{minipage}
        \begin{minipage}{0.4\textwidth}
        \centering
            \includegraphics[clip, trim = 12.1cm 0cm 12.8cm 0cm, height = 8cm]{plot/Slide5.pdf}
            \subcaption{Multiple biomarkers}
        \end{minipage}
        \caption{Extensions of our procedure include: (a) cases where the treatment effect first increases and then decreases with the biomarker, and (b) settings with multiple biomarkers where the treatment effect depends monotonically on each biomarker.}
        \label{fig:extension}
\end{figure}

We extend our procedure beyond the case where the treatment effect is monotonically related to a single biomarker.
\begin{itemize}
    \item Arc-shape dependence (\Cref{fig:extension} (a)).     
    In practice, it is possible that the effect of a new surgery might be stronger for older individuals but less effective, or even harmful, for very elderly people. 
    In this scenario, the treatment effect first increases and then decreases (arc-shape) as the biomarker value rises. 
    For this arc-shape dependence, we aim to find two thresholds $\hat{s}_- < \hat{s}_+$, and 
    select the subgroup $\calS_{\hat{s}_-, \hat{s}_+} = \{i: {\hat{s}_-} < \biomarker_i < {\hat{s}_+}\}$. 
    We can apply the method in \Cref{sec:method} to $\biomarker_i$ and the sign-flipped $-\biomarker_i$ to obtain $\hat{s}_-$ and $\hat{s}_+$, respectively. 
    The inference method for $\calS_{\hat{s}}$ remains valid for $\calS_{\hat{s}_-, \hat{s}_+}$.

    \item Multiple biomarkers (\Cref{fig:extension} (b)).
    There may exist multiple biomarkers $\biomarker_{j \cdot}$, $j \in [J]$, $J \ge 2$ influencing a drug's effectiveness, for example, the cholesterol level and BMI can collectively impact how effectively a medication lowers blood pressure. If this is the case, we can apply the method individually to each biomarker, resulting in multiple selected subgroups $\calS_{\hat{s}_{j}}$, and take the intersection $\cap_{j} \calS_{\hat{s}_{j}}$. The inference method for $\calS_{\hat{s}}$ remains valid when applied to $\cap_{j} \calS_{\hat{s}_{j}}$.

\end{itemize}

\subsection{Data-driven biomarkers}

\begin{figure}[tbp]
        \centering
        \begin{minipage}{0.4\textwidth}
        \centering
            \includegraphics[clip, trim = 22.2cm 0cm 0cm 0cm, height = 8cm]{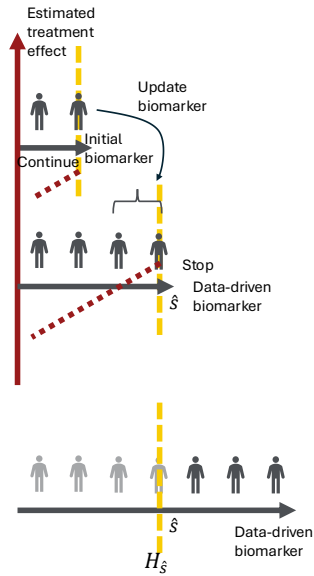}
        \end{minipage}
        \caption{Extension of our procedure to data-driven biomarkers.}
        \label{fig:extension.adaptive}
\end{figure}

If no natural biomarker is available, algorithms can be employed to construct a score, such as the genetic score used in the analysis of the FOURIER (Further Cardiovascular Outcomes Research With PCSK9 Inhibition in Subjects With Elevated Risk) trial \parencite{marston_predicting_2020}. More broadly, any individual treatment effect (HTE) predictor can serve as an artificial biomarker.

In \Cref{fig:extension.adaptive}, we demonstrate how the test described in \Cref{sec:method} can be integrated with a biomarker learned from the same dataset. 
In particular, we use an estimated CATE as the biomarker.
Explicitly, for initialization, a small subset of the data, referred to the training set, is revealed, and a heterogeneous treatment effect estimator is trained to serve as the initial data-driven biomarker. 
Subsequently, we order the remaining units not yet utilized according to their predicted treatment effect.
We reveal the outcomes of the next unit or batch with the minimal predicted treatment effect, and incorporate the data into the training dataset.
If the estimated effect of the current batch is significantly positive, or satisfies a certain criterion, the process is stopped and the units additional to the training is output.
Otherwise, we update the data-driven biomarker based on the enlarged training dataset.
The updated biomarker is used to reorder the remaining units. The process continues with revealing the next batch and reevaluating the stopping criterion.
The iterative approach progressively incorporates more data points and refines the biomarker as well as the ordering of the units. 
The details are summarized in \Cref{algo:adaptive}.

\begin{algorithm}
\caption{Subgroup selection with data-driven biomarker and selective randomization inference}
\label{algo:adaptive}
\begin{algorithmic}
\STATE
\textbf{Input}: Data $(X_i, Z_i, Y_i)$ ($X_i$ covariates), test statistic function $T(z, w)$, heterogeneous treatment effect estimation method, treatment assignment mechanism $\pi$, number of permutations $M$.
\STATE \textbf{Initialization}. Randomly sample a small subset of the data as the training dataset (denoted by $\calT$). Denote the rest of the units by $\calS$. 
\STATE \textbf{Stage 1: Subgroup selection.} 

\WHILE{True}
\STATE Apply the heterogeneous treatment effect estimation method to $\calT$. Denote the predicted treatment effect for $i \in \calS$ by $\hat{\biomarker}_i$.\\
\STATE Ordering. Order units in ascending order of $\hat{\biomarker}_i$ (randomly breaking ties if necessary). 
\STATE Reveal the next batch with the minimal predicted treatment effect $\hat{\biomarker}_i$.
Add the next batch to \(\calT\) and remove it from \(\calS\).
\IF{Stopping criterion of the next batch is met}
\STATE Break.
\ENDIF
\ENDWHILE
\STATE \textbf{Stage 2: Selective randomization inference.} 

Impute potential outcomes under $H_{\calS}$. 
\FOR{t $= 1: M$}
\STATE Generate a new proposal $\treatment^{(t)}$ of the selected subgroup $\calS$ following the conditional distribution $\pi(\cdot \mid \treatment_{\calS^c}^* = \treatment_{\calS^c})$.
\ENDFOR
\STATE Compute $\hat{P}_M = \left({1+\sum_{t=1}^{M} \one\{T(\treatment^{(t)}, \dataConditionedOn) \leq T(\treatment, \dataConditionedOn)\}}\right)/({1+M})$. 
\STATE \textbf{Output}: $\calS$, $\hat{P}_M$.
\end{algorithmic}
\end{algorithm}

Since the data-driven biomarker is trained on the complement of the final selection set \( \mathcal{S} \), the following type I error control follows directly as a corollary of \Cref{prop:typeIError}. 
\begin{corollary}\label{coro:typeIError.adaptive}
    The randomization p-value for the selected subgroup $\calS$ in \Cref{algo:adaptive}, controls the conditional type I error,
    \begin{align*}
        \Prob\left(\,P(Z, W) \leq \alpha \,\vert\, W, \calS\right) \leq \alpha,\quad \forall\, ~ \alpha \in [0,1].
    \end{align*}
    and the associated marginal type I error.
\end{corollary}

For power, the effectiveness of the selection largely depends on the quality of the CATE estimator. Our simulations show that, compared to sample splitting, which sacrifices a constant proportion of the set of interest, our power is significantly higher. Additionally, the performance of the data-driven biomarker is comparable to, though slightly worse than, that of a perfectly pre-specified biomarker.

\section{Proofs}\label{appe:sec:proofs}

\begin{proof}[Proof of \Cref{prop:typeIError}]
    For any $\hat{s}$, let $t = \max\{\biomarker_i: \biomarker_i \le \hat{s}\}$, which is a stopping time adapted to the filtration $\calF_i$, and $\biomarker_{t+1} > \biomarker_{t}$ if $t < n$.
    The selected subgroups implied by $\hat{s}$ and $t$ are the same, and thus it is sufficient to prove \Cref{eq:typeIError} for threshold $\biomarker_t$.
    We rewrite the stopping time $t$ in \Cref{sec:method} as $t(z, w)$ and $t$ depends on $y(0)$, $y(1)$ through $y(z)$.

    We define a mapping
    \begin{align*}
        G_{t}(z, w) = [z_1, \ldots, z_{t(z, w)}], \quad z \in \calZ. 
    \end{align*}
    Let $\calG_{t}(w) \subseteq \cup_{t=1}^n \{0,1\}^t$ be all possible values of $G_{t}(z, w)$ for $z \in \calZ$. 
    Given $W$, $\cup_{g \in \calG(W)} \Sigma_g(W) := \{z: G_{t}(z, W) = g\}$ forms a countable partition of $\calZ$. 
    We show a stronger conditional type I error control
    \begin{align}\label{proof:eq:conditionalTypeIError}
        \Prob\left(\,p(Z, W) \leq \alpha \,\vert\, W, G_t(Z, W)=g\right) \leq \alpha,\quad \forall\, g\in \calG_t(W), \alpha \in [0,1].
    \end{align}
    \Cref{prop:typeIError} can be obtained from the conditional type I error \eqref{proof:eq:conditionalTypeIError} and tower property since $t(z,w)$ is a function of $G_t(z, w)$,

    To prove \eqref{proof:eq:conditionalTypeIError}, it is sufficient to show $z^*_{\calS_{\hat{s}}} = z_{\calS_{\hat{s}}}$ is equivalent to $z^* \in \Sigma_{G_{t}(z, W)}(W)$ according to \cite{zhang_2023_randomization_test}.
    In fact, On one hand, $G_{t}(z, W) = G_{t}(z^*, W)$ implies $t(z, W) = t(z^*, W)$, therefore, 
    \begin{align*}
        z_{\calS_{\hat{s}}}
        = [z_1, \ldots, z_{t(z, W)}] = G_{t}(z, W) 
        = G_{t}(z^*, W) = [z^*_1, \ldots, z^*_{t(z^*, W)}] = [z^*_1, \ldots, z^*_{t(z, W)}]
        = z^*_{\calS_{\hat{s}}}.
    \end{align*}
    For the other side, we prove by negation. Suppose $z_{\calS_{\hat{s}}} = z^*_{\calS_{\hat{s}}}$ or equivalently $[z_1, \ldots, z_{t(z, W)}] = [z^*_1, \ldots, z^*_{t(z, W)}]$, but $G_{t}(z, W) \neq G_{t}(z^*, W)$.
    Then $t(z, W) \neq t(z^*, W)$.
    If $t(z, W) < t(z^*, W)$, then $Y_i(z^*_i) = Y_i(z_i)$ for any $i \le t(z, W)$.
    Together with $t$ satisfying (a), we have $t(z^*, W) = t(z, W)$. Contradiction!
    If $t(z, W) > t(z^*, W)$, then the condition $[z_1, \ldots, z_{t(z, W)}] = [z^*_1, \ldots, z^*_{t(z, W)}]$ implies $[z_1, \ldots, z_{t(z^*, W)}] = [z^*_1, \ldots, z^*_{t(z^*, W)}]$. We can swap $z$ and $z^*$ and apply the above argument.    
\end{proof}

\begin{proof}[Proof of \Cref{prop:power}] 

Without loss of generality, we assume that the units are ranked in increasing order of their biomarker values, and that there are no ties (if ties exist, we can add an independent noise of order \( n^{-1} \) to each biomarker value to break the ties without affecting the accuracy of the estimated zero point).
Let $\calO$ denote the collection of the observed covariates and the biomarker value.
Let \( e_i \) denote the propensity score of the \( i \)th unit. Without loss of generality, we assume that both \( n^{2/3} \) and \( n^{1/3} \) are integers. 
Define \( \mathcal{B}_j \), for \( 1 \le j \le n^{1/3} \), as the \( j \)th bin containing the \(((j-1)n^{2/3} + 1 )\)st unit to the \( (j n^{2/3})\)th unit, with breakpoints at the \( (j n^{2/3}) \)th unit.
Let \( j^* \) denote the index of the bin that contains the true zero point $s^*$ of $\tau(s)$.

For bin $\calB_j$, the estimator of its ATE takes the form
\begin{align*}
    \hat{\tau}_j 
    := \frac{\sum_{i \in \calB_j} Y_i Z_i }{\sum_{i \in \calB_j} e_i}
    - \frac{\sum_{i \in \calB_j} Y_i (1-Z_i) }{\sum_{i \in \calB_j} (1-e_i) }
    = \frac{1}{n^{2/3}} \sum_{i \in \calB_j} \left(\frac{Y_i Z_i }{\sum_{i \in \calB_j} e_i/n^{2/3}}
    - \frac{Y_i (1-Z_i) }{\sum_{i \in \calB_j} (1-e_i)/n^{2/3} } \right).
\end{align*}
Condition on $\calO$, $\hat{\tau}_j$ is the avearge of a set of independent sums, is unbiased of $\tau_j := \sum_{i \in \calB_j}\tau(S_i)$.
By the overlap assumption, $\sum_{i \in \calB_j} e_i \ge \sum_{i \in \calB_j} \underline{e} = n^{2/3} \underline{e} $, $\sum_{i \in \calB_j} (1-e_i) \ge \sum_{i \in \calB_j} (1-\underline{e}) = n^{2/3} \underline{e}$.
Therefore, 
\begin{align*}
    \left|\frac{Y_i Z_i }{\sum_{i \in \calB_j} e_i/n^{2/3}}
    - \frac{Y_i (1-Z_i) }{\sum_{i \in \calB_j} (1-e_i)/n^{2/3} }\right|
    \le \frac{B}{\underline{e}}.
\end{align*}

For bin \( \mathcal{B}_j \), the estimator of its ATE (local ATE) is defined as
\begin{align*}
    \hat{\tau}_j 
    := \frac{\sum_{i \in \mathcal{B}_j} Y_i Z_i}{\sum_{i \in \mathcal{B}_j} e_i}
    - \frac{\sum_{i \in \mathcal{B}_j} Y_i (1 - Z_i)}{\sum_{i \in \mathcal{B}_j} (1 - e_i)}
    = \frac{1}{n^{2/3}} \sum_{i \in \mathcal{B}_j} \left( 
        \frac{Y_i Z_i}{\sum_{i \in \mathcal{B}_j} e_i / n^{2/3}} 
        - \frac{Y_i (1 - Z_i)}{\sum_{i \in \mathcal{B}_j} (1 - e_i) / n^{2/3}} 
    \right).
\end{align*}
Under the super-population model, conditioning on \( \mathcal{O} \), \( \hat{\tau}_j \) is the average of $n^{2/3}$ independent terms, and is an unbiased estimator of
\[
\tau_j := \sum_{i \in \mathcal{B}_j} \tau(S_i).
\]
By the overlap assumption, we have
\[
\sum_{i \in \mathcal{B}_j} e_i / n^{2/3}, ~
\sum_{i \in \mathcal{B}_j} (1 - e_i) / n^{2/3} \geq \underline{e}.
\]
Together with the assumption of bounded potential outcomes, for each \( i \in \mathcal{B}_j \), the associated summand can be bounded as
\begin{align*}
       \left|\frac{1}{n^{2/3}} \left(\frac{Y_i Z_i}{\sum_{i \in \mathcal{B}_j} e_i / n^{2/3}} 
        - \frac{Y_i (1 - Z_i)}{\sum_{i \in \mathcal{B}_j} (1 - e_i) / n^{2/3}} 
   \right) \right| 
    \leq \frac{1}{n^{2/3}} \cdot \frac{B}{\underline{e}}.
\end{align*}
By Hoeffding's concentration inequality, we have
\begin{align}\label{proof:eq:concentration}
    \begin{split}
         \PP\left(\left|\hat{\tau}_{j} - \tau_j\right| > (B/\underline{e})\sqrt{\frac{\log(n)}{n^{2/3}}} ~\big |~ \calO \right)
        \le 2 \exp\left\{-\frac{2(B/\underline{e})^2 \log(n)/n^{2/3}}{4 n^{2/3} \cdot (B / n^{2/3} / \underline{e})^2}\right\}
        =  2 \exp\left\{-\frac{\log(n)}{2}\right\}
        = \frac{2}{\sqrt{n}}.
    \end{split}
    \end{align}
    Define $\calD_j := \{\left|\hat{\tau}_{j} - \tau_j\right| > (B / \underline{e}) \sqrt{\log(n)/n^{2/3}}\} $, and let $\calD = \cup_{1 \le j \le n^{1/3}} \calD_j$, then by Eq.~\eqref{proof:eq:concentration} and the union bound,
    \begin{align*}
        \PP\left(\calD\right)
        \le \sum_{1 \le j \le n^{1/3}} \PP(\calD_j)
        = \le \sum_{1 \le j \le n^{1/3}} \EE[\PP(\calD_j \mid \calO)]
        \le n^{1/3} \cdot \frac{2}{\sqrt{n}}
        = 2 n^{-1/6}.
    \end{align*}

    We first show the algorithm will not significant undershoot (stop early). 
    For \( \mathcal{B}_j \) whose largest point is at least $(1/ \underline{L})(B / \underline{e}) \sqrt{\log(n) / n^{2/3}}$ below the zero point, then by the Lipschitz condition,
    \begin{align*}
        \tau_j \le - \min \left\{\delta, ~\underline{L} \cdot (1/ \underline{L})(B / \underline{e}) \sqrt{\log(n) / n^{2/3}} \right\}
        \le -(B / \underline{e}) \sqrt{\log(n) / n^{2/3}}
    \end{align*}
    for $n$ large enough.
    Therefore, on the event $\calD$,  $\hat{\tau}_j < 0$ and the algorithm will not stop in $\calB_j$.

    Next, we show that the algorithm won't overshoot (stop late).
   Define the interval $\calI := [z^* + (B/\underline{e})\sqrt{\log(n)/n^{2/3}} / (\underline{f} \underline{L}), z^* + 2 (B/\underline{e})\sqrt{\log(n)/n^{2/3}} / (\underline{f} \underline{L})]$. 
     Let $V_i = \1_{\{s_i \in \calI\}}$.
     Then since the biomarker has density lower bounded by $\underline{f}$ in a neighborhood of $s^*$, we have $\EE[V_i]/n \ge (B/\underline{e}) \sqrt{\log(n)/n^{2/3}} / \underline{L}$ for $n$ large enough.
     By Hoeffding's concentration inequality, 
    \begin{align*}
        \PP\left(\sum_{1 \le i \le n} V_i \le 2 n^{2/3}\right)
        \le  \PP\left(\sum_{1 \le i \le n} (V_i -\EE[V_i])\le  n^{2/3} \left(2-(B/\underline{e}) \sqrt{\log(n)} / \underline{L}\right)\right)
        \le \exp\left\{-C n^{1/3}\right\}
    \end{align*}
    for some $C > 0$.
    As a result, there is at least one bin $\calB_{j^\calI} \subseteq \calI$.
    By the assumption of the lower bound of the Lipschitz constant, the $\tau_{j^\calI}$ therein is at least $(B / \underline{e}) \sqrt{\log(n) / n^{2/3}}$.
    Then under $\calD$, $\hat{\tau_{j^\calI}} > 0$, and the algorithm must stop before or at $B_{j^\calI}$.

    Finally, we combine the no overshooting and no undershooting cases and complete the proof.

\end{proof}

\section{Additional simulations}\label{appe:sec:simulation}

\begin{figure}[tbp]
        \centering
        \begin{minipage}{0.35\textwidth}
        \centering
    \includegraphics[clip, trim = 0cm 0cm 6cm 0cm, height = 0.9\textwidth]{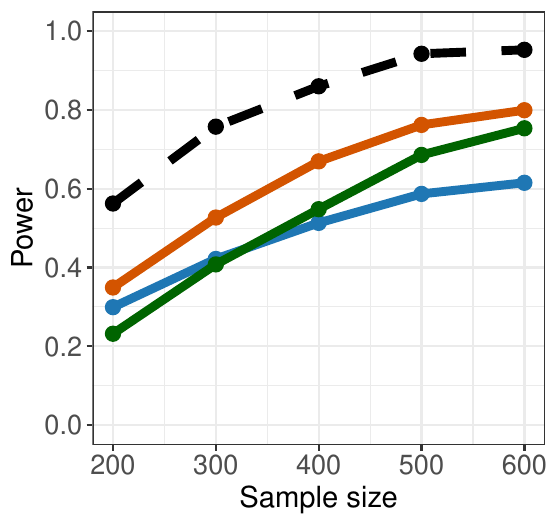}
        \end{minipage}
       \hspace{0.5cm}
        \begin{minipage}{0.35\textwidth}
        \centering
            \includegraphics[clip, trim = 0cm 0cm 0cm 0cm, height = 0.9\textwidth]{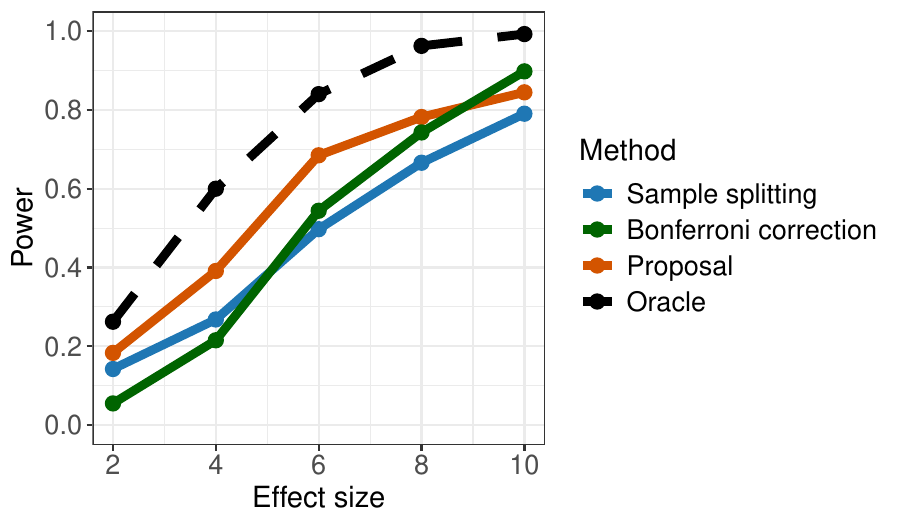}
        \end{minipage} 
\caption{Power comparison of four randomization-based methods. 
}
        \label{fig:power.piece.wise.constant}
\end{figure}

We include the comparison of the four methods in \Cref{sec:simulation} under a piecewise constant $\tau$. 
As illustrated in \Cref{fig:power.piece.wise.constant}, the power comparison is similar to those in \Cref{sec:simulation}. 
We note that the gap between our proposal and the oracle method widens in this setting, as a smaller bin or narrower bandwidth is needed in \Cref{algo:1} to capture the abrupt change in the treatment effect.

\section{Additional tables}\label{appe:sec:table}

We provide a summary table of the treatment and control group characteristics from the GBCS dataset.

\begin{table}[tbp]
\centering
\caption{Summary of covariates in the GBCS dataset. For each covariate, we provide the mean and standard deviation (in the parentheses) in the treatment group and the control group, respectively.}
\label{tab:data}
\begin{tabular}{cc|cc|cc}
\toprule
Covariate & Description & \multicolumn{2}{c|}{Treatment group}                     & \multicolumn{2}{c}{Control group}      \\ 
& & \multicolumn{2}{c|}{(n=178)}                     & \multicolumn{2}{c}{ (n=261)}      \\ \midrule
Age   & Age at diagnosis (years).             & 55.25          & (9.93)     & 51.27        & (10.44)      \\
Menopause & Menopausal status: 1 = Yes, 0 = No.           & 0.69        & (0.47)      & 0.48       & (0.50)      \\
Size  &  Tumor size (mm).           & 28.75          & (14.75)       & 29.26         & (15.04)      \\
Grade   & Tumor grade (1-3).             & 2.08       & (0.57)      & 2.14         & (0.59)     \\
Nodes      & Number of nodes.          & 5.02         & (4.98)      & 4.72         & (5.03)     \\
$\log$(Prog recp)  & Log of number of progesterone receptors.    & 3.23        & (2.01)      & 3.30         & (1.94)      \\
$\log$(Estrg recp)  & Log of number of estrogen receptors. & 3.45       & (1.94)       & 3.18 & (1.87) \\ \bottomrule
\end{tabular}
\end{table}

\end{document}